# One and two proton separation energies from nuclear mass systematics using neural networks

S. Athanassopoulos,[a] E. Mavrommatis,[a] K. A. Gernoth,[b] J. W. Clark[c]

[a]*Physics Department, Division of Nuclear & Particle Physics, University of Athens, GR-15771 Athens, Greece*

[b]*Department of Physics, UMIST, P.O.Box 88, Manchester M60 1QD, United Kingdom*

[c]*McDonnell Center for the Space Sciences and Department of Physics, Washington University, St. Louis, Missouri 63130, USA*

**Abstract**

We deal with the systematics of one and two proton separation energies as predicted by our latest global model for the masses of nuclides developed with the use of neural networks. Among others, such systematics is useful as input to the astrophysical rp-process and to the one and two proton radioactive studies. Our results are compared with the experimental separation energies referred to in the 2003 Atomic Mass Evaluation and with those evaluated from theoretical models for the masses of nuclides, like the FRDM of Möller et al. and the HFB2 of Pearson et al. We focus in particular on the proton separation energies for nuclides that are involved in the rp-process ($29 \leq Z \leq 40$) but they have not yet been studied experimentally.

## 1. Introduction

In this work we present global models for the one proton and two proton separation energies of nuclei, defined respectively in terms of the binding energies $B$ or mass excesses $\Delta M$ as follows:

$$S(p) = B(A,Z) - B(A-1, Z-1) = \Delta M(Z-1, N) + \Delta M_H - \Delta M(Z, N)$$

$$S(2p) = B(A,Z) - B(A-2, Z-2) = \Delta M(Z-2, N) + 2\Delta M_H - \Delta M(Z, N)$$

The problem of devising global models of the proton separation energies is mostly connected with the problem of devising global models for the atomic masses or binding energies of nuclides. Besides providing an understanding of the physics of the mass (binding energy) surface they are useful for prediction of these properties for "new" nuclides far from stability. These predictions are of current interest in connection with the experimental studies of nuclei far from stability conducted at heavy-ion and radioactive ion-beam facilities as well as for such astrophysical problems such as nucleosynthesis and supernova explosions [1]. In particular, the global models of the proton separation energies are useful mainly in the study of proton and two proton radioactivity [2] and the rp-process of nucleosynthesis [3]. The latter is believed to take place on the surface of white dwarfs (novae) and of neutron stars (type I X-ray bursts), on accretion disks around low mass black holes as well as in Thorne-Zytkow objects. The rp-process may also be responsible for the p-process nucleosynthesis of a few proton-rich stable nuclei in the $A$=74-98 mass range.

The global models of the proton separation energies developed so far are mainly derived from the known global models of the atomic mass. The spectrum of the latter ranges from those with high theoretical input that take explicit account of known physical principles in terms of a relatively small number of fitting parameters to models that are shaped mostly by the data and very little by the theory and thus have a correspondingly large number of adjustable parameters. Current models of the former class that set the state of the art are the finite range droplet model (FRDM) of Möller, Nix and coworkers detailed in Refs [4,5] and the Hartree-Fock-Bogoliubov model (HFB2) of Pearson, Tondeur and coworkers detailed in Ref. [6]. There are also "restricted" global models of proton separation energies that address in detail the evaluation of proton separation energies in certain region of nuclides like the sd shell or the fp shell [7,8,9] and the suburanium and superheavy regions [10].

We use neural networks to develop global models for the one proton and two proton separation energies. In this work our models are based on our best neural network global mass model detailed in Ref. [11]. The models derived by means of neural network methodology are situated far toward the other end of the spectrum mentioned above, where one (in the ideal) seeks to determine the degree to which the entire mass table of a given property is determined by the existing data and only by the data. During the last decade artificial neural networks have been utilized to construct predictive statistical models in a variety of scientific problems ranging from astronomy to experimental high-energy to protein structure [12]. To date, global neural network models have been developed for the stability/instability dichotomy, for the atomic mass table, for neutron separation energies, for spins and parities and for decay branching probabilities of nuclear ground states and for $\beta^-$ decay half-lives [11,13].

In a typical example, a multilayered feed-forward neural network is trained with a supervised training algorithm to create a "predictive" statistical model of a certain input-output mapping. Information contained in a set of learning examples of the input-output association is embedded in the weights of the connections between the layered units in such a way that the network provides a means for interpolation or extrapolation.

In section 2 we outline the neural network model specifications along with the data sets used for training, evaluation of predictive performance and prediction. In section 3 we summarize the results for the mass excess while in section 4 we present the corresponding results for one and two proton separation energies and we estimate the position of the proton drip line for nuclei with 29$\leq Z \leq$40. Finally, section 5 states the general conclusions of the current study and views the prospects for further improvements in statistical prediction of proton separation energies.

## 2. Neural network mass model

After a substantial number of attempts (see Ref. [11] for details) a multilayered feed-forward neural network is adopted with gross architecture summarized in the notation (4-10-10-10-1)[363]. The four units of the input layer encode the atomic number $Z$, neutron number $N$ and their respective parities, while the single unit of the output layer encodes the mass excess $\Delta M$. A scaling recipe has been used for the $Z$, $N$ and $\Delta M$ variables which allows for the ranges [0,130], [0,200] and [-110,250] respectively. The total number of weight/bias parameters that connect the input to the output layer through the three intermediate layers (each consisted of 10 units) is 363.

The training of the neural network was simple: when the training patterns were presented to the input interface, the states of all units within a given layer were updated successively, proceeding from input to output. Based on the deviation between the target and output mass excess values, the weight parameters were continuously readjusted through a minimization training algorithm. Specifically, a novel back-propagation algorithm has been used that helps to avoid the local minima during the training process. In addition, several other techniques have been used to improve training and predictive performance.

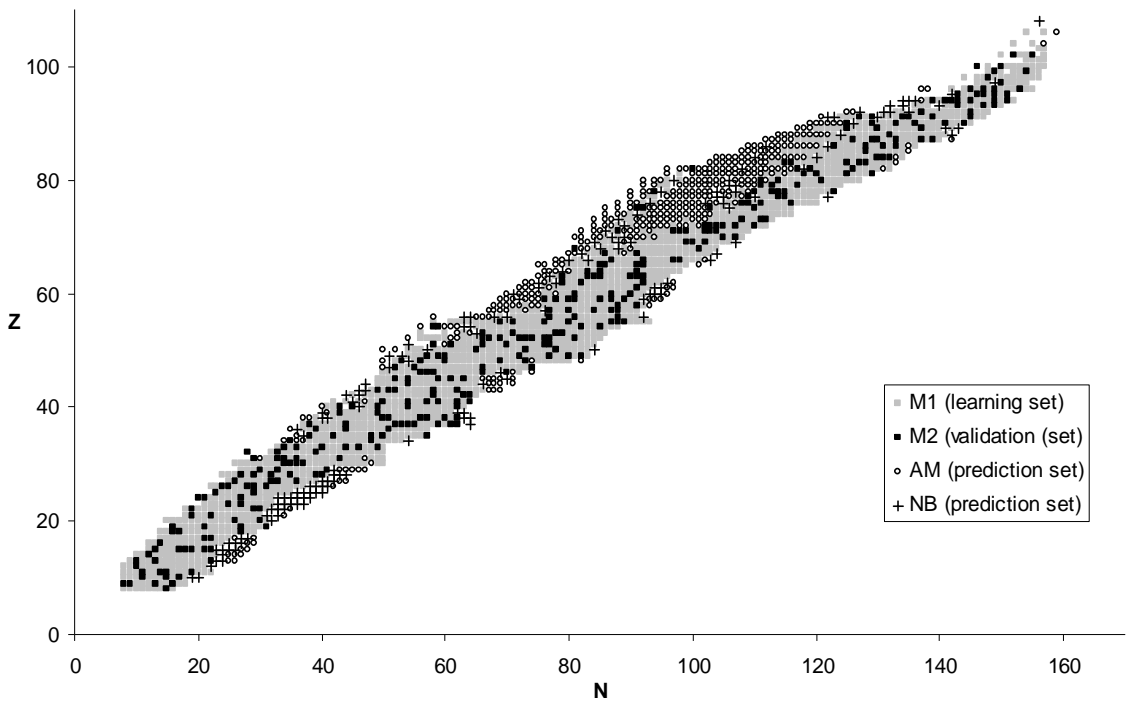

Fig. 1: Locations in the $N$–$Z$ plane are indicated for the M1, M2, NB and AM data sets employed in neural-network modeling of nuclear mass excesses.

For training the neural network mass model we have employed a database of 1654 nuclei which form the database fitted by the FRDM parameterization of Ref. [4]. We have split them randomly into two data sets of 1303 (M1) and 351(M2) nuclei that are utilized as learning and validation sets respectively. The former has being used during training for adjusting the weight parameters while the performance on the latter was used as a criterion for when to stop the training process. While the members of the validation set are not used in the weight updates, they clearly do affect the choice of model. To obtain a clean measure of predictive performance, a prediction set is needed that is never referred to during the training process. Such a set (denoted NB) was formed from 158 new nuclei drawn from the NUBASE evaluation of nuclear and decay properties [14], which lie beyond the 1654 nuclide set as viewed in the $N-Z$ plane (see Fig. 1).

**3. Mass Excess evaluation**

As performance measure we chose the root mean square error $\sigma_{RMS}$. We report in Table 1 its values on learning, validation and prediction sets for the neural network of Ref. [11] and for the FRDM [4] and HFB2 [6] models.

Table 1: Root mean square error ($\sigma_{RMS}$) of global models for the atomic mass table (see text for details).

| Model | Learning set (M1) $\sigma_{RMS}$ (MeV) | Validation set (M2) $\sigma_{RMS}$ (MeV) | Prediction set (NB) $\sigma_{RMS}$ (MeV) |
|---|---|---|---|
| FRDM (Ref. [4]) | 0.68 | 0.71 | 0.70 |
| HFB2 (Ref. [6]) | 0.67 | 0.68 | 0.73 |
| (4 – 10 – 10 – 10 –1) (Ref. [11]) | 0.44 | 0.44 | 0.95 |

Further information on the predictive performance of the neural network mass model is furnished in Figs 2 and 3. In Fig. 2, we compare the deviations from experimental data of the mass excess values generated by the net and by the FRDM evaluation, for the NB nuclei. The extrapolation capability of the neural network model is better illustrated in the Fig. 3 which shows these deviations as a function of the number of neutrons away from the $\beta$-stability line.

After completing the training of the above neural network model, the 2003 Atomic Mass Evaluation (AME03) was published [15]. This compilation made available precision mass measurements for nuclei farther off the stability line, while providing corrected mass-excess values for nuclei already used in our study. The next generation of neural-network models will be trained using the AME03 data. Already however, we can further appraise the extrapability performance of our network by making use of 376 new nuclei included in the AME03, which extend mostly beyond the edges of the M1+M2+NB nuclide set as viewed in the $N-Z$ plane. The resulting value of $\sigma_{RMS}$ for this set of nuclei (denoted AM, see Fig. 1) is 1.06 MeV, which is to be compared with the figures 0.52 MeV and 0.68 MeV obtained in the FRDM and HFB2 evaluations respectively. When comparing these results, it should be kept in mind that the parameters of the HBF2 model have been adjusted by making use of an extended data set of 1888 nuclei, which includes 102 of the 376 nuclides.

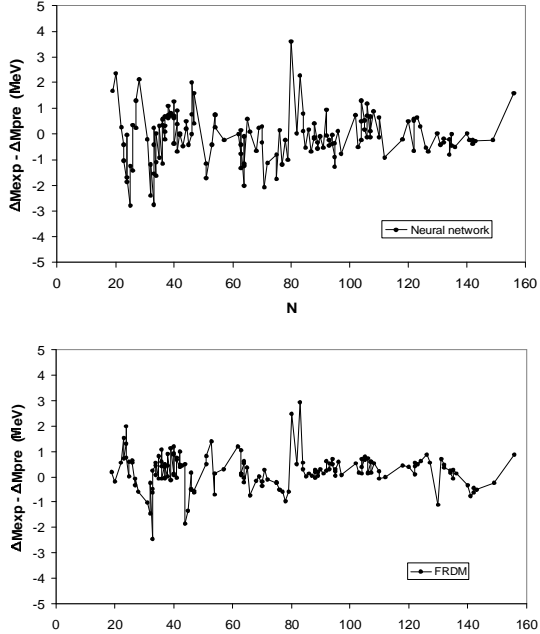
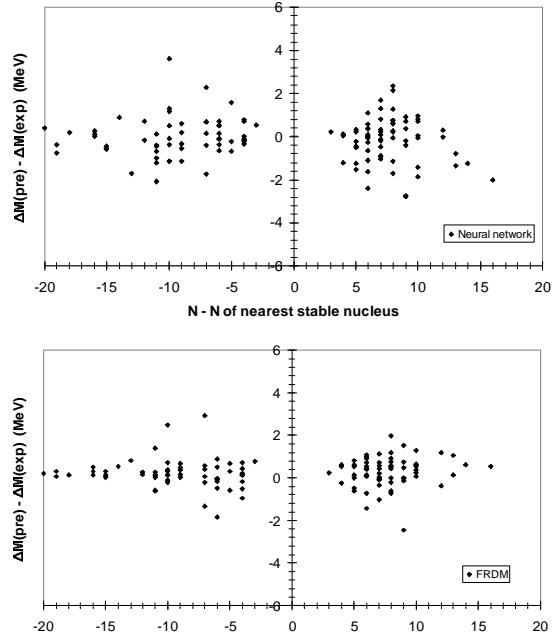

Fig. 2: Top panel: deviations from experiment (in MeV) of mass-excesses values predicted by the neural-network model [11] for the NB nuclei. The plot represents a projection of the mass surface onto a plane of constant $Z$ and thus shows dependence on neutron number $N$. Bottom panel: same for the FRDM evaluation [4].

Fig. 3: Top panel: deviations from experiment (in MeV) of mass-excesses values predicted by the neural-network model [11] for the NB nuclei plotted versus the number of neutrons away from the line of $\beta$-stability. Bottom panel: same for the FRDM evaluation [4].

## 4. Proton separation energies

The AME03 basis contains 2040 nuclei with experimentally measured one proton separation energies S(p). For 1968 of these nuclei with $Z, N \geq 8$, values for S(p) can be evaluated from the mass excess values evaluated by the FRDM, HFB2 and neural network models discussed in section 3. The corresponding $\sigma_{RMS}$ are reported in Table 2. To estimate the predictive performance of the models we report separately results for 123 and 330 nuclei of the NB and AM data sets respectively, for which one proton separation energies can also be evaluated.

The AME03 basis also contains 1900 nuclei with experimentally measured two proton separation energies S(2p). For 1846 of these nuclei with $Z, N \geq 8$, values for S(2p) can be evaluated from the mass excess values evaluated by the FRDM, HFB2 and neural network models discussed in section 3. The corresponding $\sigma_{RMS}$ are reported in Table 3. To estimate the predictive performance of the models we report separately results for 107 and 327 nuclei of the NB and AM data sets respectively, for which two proton separation energies can also be evaluated.

Table 2: Performance measures of global models of one proton separation energies derived from global models of the mass excess (see text for details).

| Model | 1968 of 2040 $\sigma_{RMS}$ (MeV) | 123 of 158 (NB) $\sigma_{RMS}$ (MeV) | 330 of 376 (AM) $\sigma_{RMS}$ (MeV) |
|---|---|---|---|
| FRDM (Ref. [4]) | 0.40 | 0.48 | 0.37 |
| HFB2 (Ref. [6]) | 0.49 | 0.49 | 0.43 |
| (4 – 10 – 10 – 10 –1) (Ref. [11]) | 0.53 | 0.72 | 0.62 |

Table 3: Performance measures of global models of two proton separation energies derived from global models of the mass excess (see text for details).

| Model | 1836 of 1900 $\sigma_{RMS}$ (MeV) | 107 of 158 (NB) $\sigma_{RMS}$ (MeV) | 327 of 376 (AM) $\sigma_{RMS}$ (MeV) |
|---|---|---|---|
| FRDM (Ref. [4]) | 0.49 | 0.54 | 0.33 |
| HFB2 (Ref. [6]) | 0.51 | 0.66 | 0.43 |
| (4 – 10 – 10 – 10 –1) (Ref. [11]) | 0.61 | 0.74 | 0.69 |

From the $\sigma_{RMS}$ values reported for S(p) and S(2p) in Tables 2 and 3, we see that the neural network models have reached extrapability levels comparable with those reached by the best global models rooted in quantum theory. The ultimate test of any global model is the accuracy that can be realized in the prediction of separation energies of nuclear species prior to measurement. It is particularly important to predict for each Z the first isotope with negative S(p) or S(2p), indicating the position of the proton drip line. For the elements with 29≤Z≤40 the position of the proton drip line has been estimated and drawn in Fig. 4 using the systematics of S(p) and/or S(2p) created by the neural network mass model (in a few cases where no negative value was predicted for either S(p) or S(2p), the minimum value has been used instead). As it was mentioned before, this region of the nuclear chart is of great current importance in the rp-process at relatively high temperature. As expected due to pairing, the odd Z proton drip line is located substantially closer to the stability line compared to the even Z proton drip line. Our results do not differ significantly from those derived by Brown et al. and presented in Fig. 3 of Ref. [9].

**5. Conclusions – Future steps**

The current generation of neural network models of the nuclear mass excess display substantially improved performance relative to earlier attempts that use neural networks to predict masses far from the valley of β stability. We have used such models to create statistical models for the one and two proton separation energies. The results suggest that with further development this approach may provide a valuable complement to conventional global models. Strong impetus for such improvement comes from studies of nucleosynthesis (especially the rp-process), proton-rich nuclei and two-proton emission.

We are currently exploring and implementing a number of refinements of neural-network approaches to the mass problem which we will use afterwards for modelling the separation energies. These include the introduction of self-constructed neural networks that will tackle the subtle regularities of the nuclear mass systematics. We have also made some initial attempts to construct a neural network model of the differences between the experimental mass-excess values and the theoretical ones given by the FRDM model [4]. Furthermore, we have made some attempts to create directly global models for separation energies with the use of neural networks trained with the experimental separation energies.

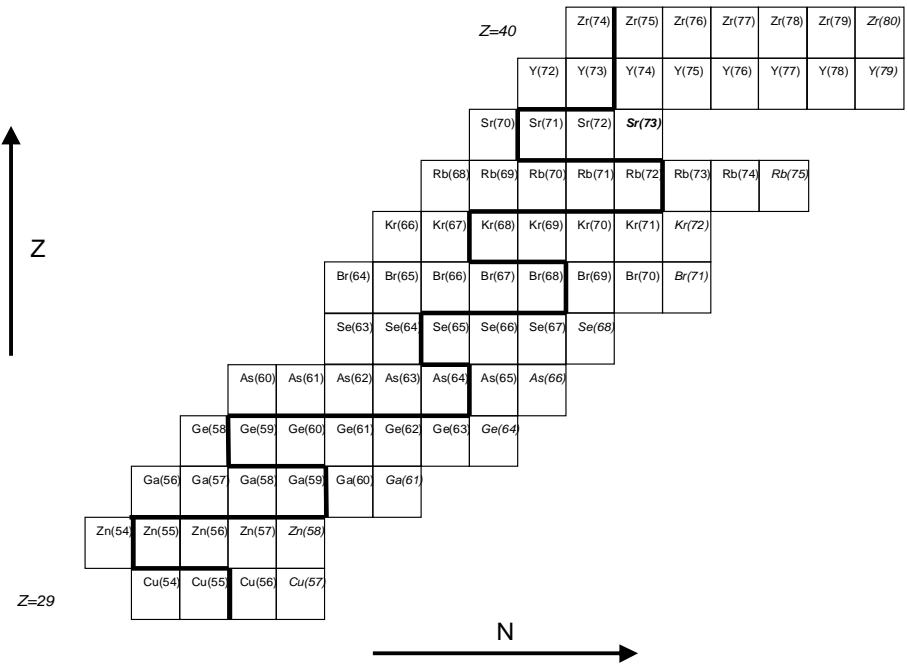

Fig. 4: Proton drip line evaluation based on the prediction of one and two proton separation energies. In italics, nuclei with experimentally measured separation energy.

**Acknowledgments**

This research has been supported in part by the University of Athens under Grant No. 70/4/3309 and by the U.S. National Science Foundation under Grant No. PHY-0140316.